\input jytex.tex   
\typesize=10pt \magnification=1200 \baselineskip17truept
\hsize=6truein\vsize=8.5truein
\footnotenumstyle{arabic}
\sectionnumstyle{blank}
\chapternumstyle{blank}
\chapternum=1
\sectionnum=1
\pagenum=0

\def\begintitle{\pagenumstyle{blank}\parindent=0pt\begin{narrow}[0.4in]}
\def\endtitle{\end{narrow}\newpage\pagenumstyle{arabic}}


\def\beginexercise{\vskip 20truept\parindent=0pt\begin{narrow}[10
truept]}
\def\endexercise{\vskip 10truept\end{narrow}}


\def\eql#1{\eqno\eqnlabel{#1}}
\def\ref{\reference}
\def\peq{\puteqn}

\def\mgn{\marginnote}
\def\bex{\begin{exercise}}
\def\eex{\end{exercise}}


 
\def\StretchRtArr#1{{\count255=0\loop\relbar\joinrel\advance\count255 by1
\ifnum\count255<#1\repeat\rightarrow}}
\def\StretchLRtArr#1{\,{\leftarrow\!\!\count255=0\loop\relbar\joinrel\advance
\count255 by1\ifnum\count255<#1\repeat\rightarrow\,\,}}
\def\mbox#1{{\leavevmode\hbox{#1}}}

\def\hspace#1{{\phantom{\mbox#1}}}


\def\al{\alpha}
\def\be{\beta}
\def\ga{\gamma}
\def\de{\delta}
\def\Ga{\Gamma}

\def\ep{\epsilon}

\def\la{\lambda}

\def\Si{\Sigma}

\def\caL{{\cal L}}


\def\frac#1/#2{\leavevmode\kern.1em
\raise.5ex\hbox{\the\scriptfont0 #1}\kern-.1em/\kern-.15em
\lower.25ex\hbox{\the\scriptfont0 #2}}
\def\sfrac#1/#2{\leavevmode\kern.1em
\raise.5ex\hbox{\the\scriptscriptfont0 #1}\kern-.1em/\kern-.15em
\lower.25ex\hbox{\the\scriptscriptfont0 #2}}

\def\gtorder{\mathrel{\raise.3ex\hbox{$>$}\mkern-14mu
             \lower0.6ex\hbox{$\sim$}}}
\def\ltorder{\mathrel{\raise.3ex\hbox{$<$}\mkern-14mu
             \lower0.6ex\hbox{$\sim$}}}

\def\semidirprod{\rlap{\ss C}\raise1pt\hbox{$\mkern.75mu\times$}}
\def\for{\lower6pt\hbox{$\Big|$}}
\def\fish{\kern-.25em{\phantom{abcde}\over \phantom{abcde}}\kern-.25em}



\def\CompressMatrices{\ifmmode\def\quad{\hskip.5em\relax}\fi}

\def\boxit#1{\vbox{\hrule\hbox{\vrule\kern3pt
        \vbox{\kern3pt#1\kern3pt}\kern3pt\vrule}\hrule}}
\def\dalemb#1#2{{\vbox{\hrule height .#2pt
        \hbox{\vrule width.#2pt height#1pt \kern#1pt
                \vrule width.#2pt}
        \hrule height.#2pt}}}

\def\frac#1#2{{{#1}\over{#2}}}

\def\noin{\noindent}


\def\eg{{\it e.g. }}
\def\ie{{\it i.e. }}
\def\cf{{\it cf }}
\def\pa{\partial}


\def\sssgap{\vskip 5truept}

\def\3j#1#2#3#4#5#6{\left\lgroup\matrix{#1&#2&#3\cr#4&#5&#6\cr}\right\rgroup}

\def\m?{\mgn{?}}

\def\pa{\partial}

\def\beq{\begin{eqnarray}}
\def\eeq{\end{eqnarray}}


\def\aop#1#2#3{{\it Ann. Phys.} {\bf {#1}} (#2) #3}

\def\pl#1#2#3{{\it Phys. Lett.} {\bf {#1}} (#2) #3}

\def\pr#1#2#3{{\it Phys. Rev.} {\bf {#1}} (#2) #3}

\def\prl#1#2#3{{\it Phys. Rev. Lett.} {\bf #1} (#2) #3}


\pagenum=0 \centertext {\Bigfonts \bf The Theory of Invariants and
Interaction} \vskip7truept \centertext{\Bigfonts\bf Symmetries} \vskip
20truept \centertext{J.S.Dowker} \sssgap \centertext{\it Department of
Theoretical Physics,} \centertext{\it The University of Manchester, Manchester, 
UK.}
 \section{\bf The background. Added 2001.}
 To judge by the recent appearance of a pedagogical textbook ({\it Classical 
Invariant
 Theory} by P.J.Olver, LMS, 1999) the constructional aspect of classical 
invariant
 theory has made a strong comeback, due, in no short measure, to the existence
 of powerful computer algebra packages and to applications in image recognition 
and
 computer aided design.

The present author was always particularly struck by the Gordan-Hilbert
finite {\it algebraic} basis theorem which was the culminating swan-song
(as it turned out) of 60 years of endeavour and, one might almost say, of
hard labour. This eminent position has not been reflected by extensive
applications in physics, where {\it linear} independence seems to rule the
everyday roost.

In 1965 I attempted to apply the theorem to an approach to particle
symmetries which has fallen by the wayside. Nevertheless, despite the naive
setup (by today's standards)  and in view of the mentioned increased
activity, I have decided to reissue the report produced at that time as I
think it still contains something of value. No changes have been made.
\newpage
\noin Not for publication
 \vglue 1truein \vskip15truept
\centertext {\Bigfonts \bf The Theory of Invariants and Interaction}
\vskip7truept \centertext{\Bigfonts\bf Symmetries} \vskip 20truept
\centertext{J.S.Dowker} \sssgap \centertext{\it Department of Theoretical
Physics,} \centertext{\it The University, Manchester.} \vskip 20truept
\vskip10truept

Recently a number of attempts have been made to derive elementary particle
interaction symmetries, including Lorentz invariance, by essentially
algebraic means. There seem to be various, more or less equivalent, ways of
deriving the basic equations. In references [1-3] bootstraps are referred
to, in references [4,5] appeal is made to the Shmushkevich principle while
in references [6,7] the basic equations are derived from the compositeness
conditions, $Z_3=0$, $Z_1=0$. However, in all derivations some
approximation is needed, whether it be a perturbation expansion or a Born
approximation. Now it turns out that only the lower perturbation
approximations are needed. In other words we appear to have more
information than is necessary and so we are led to the idea that perhaps
the lower order equations are somehow exact.

In this report we should like to introduce some ideas which suggest a
non-perturbative method for deriving coupling constant structure equations
and also, less ambitiously, indicate why only lower order perturbation
equations need to be considered. We should state at the beginning that no
definite results have yet been achieved but we feel the technique is
sufficiently of interest to merit some attention.

Since we are only interested in deriving the general structure of coupling
constant equations and not in the derivation of symmetries from these
equations we shall restrict ourselves to a system described by the
Lagrangian
  $$
  \caL={1\over2}\pa^\mu\phi^i\pa_\mu\phi^i-{1\over2}m^2\phi^i\phi^i+\caL'
  \eql{1a}$$
  where the interaction $\caL'$ is given by
  $$
  \caL'=\sum_{ijk}g_{ijk}\phi^i\phi^j\phi^k
  \eql{2a}$$
and $\phi$ is an $n$-component field and the $g_{ijk}$ are, of course,
completely symmetric.

Let us write down the `$g$--structure' equations which we are trying to
derive, [8]
  $$
  \sum_{jk}g_{ijk}\,g_{i'jk}=N\de_{ii'}
  \eql{3a}
  $$
  $$
\sum_{i'j'k'}g_{i'j'k}\,g_{ij'k'}\,g_{i'jk'}=\la\,g_{ijk}
  \eql{4a}
  $$

We now observe, with Cutkosky [2], that (\peq{3a}) and (\peq{4a}) are
covariant under orthogonal transformations $O$ \ie under the
transformations
  $$
  g_{ijk}\to\overline g_{ijk}=O_{ii'}O_{jj'}O_{kk'}\,g_{i'j'k'}
  \eql{5a}$$
where
  $$
  \widetilde OO=O\widetilde O=1
  $$
Let us now symbolically represent $g_{ijk}$ by a product of three `vector'
factors:
  $$
  g_{ijk}=\al_i\al_j\al_k
  \eql{6a}$$
so that the interaction becomes, symbolically, a perfect cube
  $$
  \caL'=(\al_i\phi^i)^3=(\tilde\al\phi)^3\equiv[\al\phi]^3
  \eql{7a}$$
Equations (3) and (4) now read on multiplying through by as many $\phi$'s
as are needed to contract all the spare indices
  $$
  [\al\phi][\be\phi][\al\be]^2=N\,[\phi\phi]
  \eql{8a}$$
and
  $$
  [\al\phi][\be\phi][\ga\phi][\al\be][\be\ga][\ga\al]=\la\,[\al\phi]^3
  \eql{9a}$$
In accordance with the usual rules [9] no more than three similar vector
symbols should occur in the same factor and this has been adhered to in (8)
and (9). These equations are orthogonally covariant under the
transformation
  $$\eqalign{
  \al\to\overline\al&=O\al\cr
  \phi\to\overline\phi&=O\phi}
  $$

There exists a considerable mathematical literature [9] concerned with
invariants and invariant (covariant) equations. On scanning through the
main results of the theory one comes across the fundamental Gordan-Hilbert
expansion theorem, which says essentially that given any fundamental
(invariant) ground form, \eg $\caL'$, then any concomitant of this ground
form can be expanded as a generalised {\it polynomial} in a {\it finite}
set of so called irreducible concomitants. A concomitant is any combination
of coefficients, \eg $g$'s, and variables, \eg $\phi$'s, which, under the
transformation in question, retains its {\it form} but is multiplied by
some power of the determinant of the transformation. In particular an
invariant is multiplied by the zeroth power, so that for orthogonal
transformations all concomitants are, in effect, invariants.

Our method can now be expressed in the above language as follows. We take
as our fundamental ground form the interaction $\caL'$ and assume it to be
an invariant under the orthogonal group. The free Lagrangian, $\caL-\caL'$
is invariant under this group, and so therefore will be the $S$--matrix. If
we expand the $S$--matrix in normal products of $\phi_0^i$, the interaction
picture fields, thus :
  $$
  S_{\rm connected}=1+\Phi_i:\phi_0^i:+\Si_{ij}:\phi_0^i\phi_0^j:+\Ga_{ijk}
  :\phi_0^i\phi_0^j\phi_0^k:+\ldots
  $$
(suppressing space-time integrations)  we can say that each term will be a
concomitant of different order (\ie power of $\phi_0$). Thus in particular
we can say that $\Si_{ij}\phi_0^i\phi_0^j$ will be expressible as a
generalised polynomial of irreducible concomitants of total order two. If
we were to write this out fully we could equate powers of $\phi_0$ and end
up with an expression for $\Si_{ij}$ in terms of combinations of the
$g_{ijk}$. The self consistency equations result now by putting
  $$
  \Si_{ij}\propto\de_{ij}
  $$
and similarly
  $$
\Ga_{ijk}\propto g_{ijk}
  $$

The problems remaining are to determine the irreducible set of concomitants
and to write down the appropriate generalised polynomial. The first problem
is a purely mathematical one of considerable complexity, [9], which becomes
virtually impenetrable for cubics (like $\caL'$) of order greater than
three (ternary cubics). Littlewood [9] has derived, more or less, the
complete set of irreducible concomitants for the orthogonal ternary cubic
using non-symbolic methods. The case of the binary cubic (\ie $\phi$--two
component) is somewhat peculiar since equations (3) and (4) yield
$g_{ijk}=0$ (unless $\la=0$)(see remarks later). It must be remarked that
Littlewood's analysis applied to the case when the ternary cubic is simple
under the orthogonal group. This means that the $g$'s are traceless \ie
$[\al\al]=0$, (This {\it follows} from equations (3) and (4) which we are
trying to derive). Presumably this restriction can be lifted using the
Gordan--Capelli expansion and the Clebsch--Young theorem [9].

However, even granted that we have a complete set at our disposal we still
have to form the generalised polynomial of order two. This appears to be
rather complicated due to the large number of irreducible concomitants (26
for the orthogonal ternary cubic). It is true that we can throw away some
combinations if they contain a single $\ep_{ijk\ldots}$ factor as such a
term cannot arise in $\Si_{ij}$ but the remainder still seems too complex
to handle easily. However, it does seem that the Gordan-Hilbert theorem
provides us in principle with a method of exhibiting the $g$--structure of
$\Si_{ij}$ and $\Ga_{ijk}$ in terms of expressions with an upper limit to
their {\it connected} complexity.

If this is true, then it is reasonable to suppose that in some way the
$g$--structure expressions occurring in the perturbation expansion are
reducible above a certain stage. This follows from the so-called
fundamental theorem [9], {\it viz} Every form which can be derived from
$g_{ijk}$, $\ep_{ijk\ldots}$ and $\phi^i$ by the usual tensor rules of
contraction is a concomitant of the ground form $\caL'$. The terms in the
perturbation expression fall into this class and since there are a finite
number of irreducible concomitants, [11], clearly most of the perturbation
terms will be reducible. This can be checked for the binary cubic by using
the determinant identity
  $$
  2(ab)(bc)(cd)(da)=(ab)^2(cd)^2+(bc)^2(da)^2-(ac)^2(bd)^2
  $$
(and extensions)  where
 $$
 (ab)=a_1b_2-a_2b_1
 $$
and
  $$
  (ab)(cd)=[ac][bd]-[ad][bc]
  $$
In this case the triangle vertex graph is reducible as can be seen by
writing out equation (4) in full. (This is not true for the ternary case.
The triangle expression occurs in the list of irreducible concomitants give
by Littlewood, [9]).

The corresponding results for the ternary case have not been derived. The
difficulty seems to be that classical invariant theory deals with
quantities like $(ab)$, $(abc)$,$\ldots$ \ie $\ep^{ij}a_ib_j$,
$\ep^{ijk}a_ib_jc_k$, $\ldots$, (We see here a link between binary
quantities and two-spinors, [10]), and all the methods of reduction are
appropriately expressed. However the perturbation series generates
concomitants in terms of scalar products $[ab]$, \ie $a_ib_i$. A
translation of the reduction process into tensor or group theory language
would seem to be necessary, [12] (\cf Littlewood [9]).
\newpage
\noin (1) Chan,H-M, P.C.De Celles and J.E.Paton, \prl{11}{1963}521

\noin(2) R.E.Cutkosky, \pr {131}{1963}{1886}; \aop{23}{1963}{415}

\noin(3) E.C.G.Sudarshan, \pl {9}{1964}{286}

\noin(4) E.C.G.Sudarshan, L.S. O'Raifeartaigh and T.S.Santhanam, \pr
{136}{1964}{B1092}

\noin(5) E.C.G.Sudarshan, {\it Symmetry in Particle Physics}, APS, Chicago
(1964)

\noin(6) J.S.Dowker, Nuovo Cimento {\bf34} (1964) {773}

\noin(7) J.S.Dowker and P.A.Cook, Nuovo Cimento, submitted 1965.

\noin(8) J.S.Dowker and J.E.Paton, Nuovo Cimento {\bf 30} (1963) {540}

\noin(9) \eg J.H.Grace and A.Young, {\it Algebra of Invariants}, Cambridge
(1903)

H.W.Turnbull, {\it Theory of Determinants}, Dover (New York 1960)

E.B.Elliot, {\it The Algebra of Quantics}, Oxford (1908)

R.Weitzenb\"ock, {\it Invariantentheorie}, Groningen (1923)

D.E.Littlewood, Phil. Trans. Roy. Soc. (A) {\bf239} (1944) 305

H.Weyl, {\it Classical Groups}, Princeton (1939)

O.E.Glenn, {\it Theory of Invariants}, Boston (1915)

A.I.Mal'cev, {\it Linear Algebra}, Freeman, San Francisco (1963)

\noin(10) H.Weyl, {\it Theory of Groups and Quantum Mechanics}, Methuen, London 
.

\noin(11) The irreducible set of concomitants is {\it algebraically}
over-complete due to the existence of syzygies, \ie identities between the
concomitants, which allow one to eliminate, in a {\it non-rational} way,
some of the concomitants. The consequence of syzygies for the perturbation
series is unclear.

\noin(12) Compare the Clebsch-Gordan series.

\bye